\documentclass{ws-procs9x6}

\begin{document}

\title{Recent Results from the SIMPLE Dark Matter Search}

\author{TA Girard$^{1}$, \underline{F. Giuliani}$^{1}$, J.I. Collar$^{2}$, T. Morlat$^{1}$, D. Limagne$^{3}$, G. Waysand$^{3,4}$, M. Auguste$^{4}$, D. Boyer$^{4}$, A. Cavaillou$^{4}$, H.S. Miley$^{5}$, M. da Costa$^{6,7}$, R.C. Martins$^{7}$, J.G. Marques$^{6}$, A.R. Ramos$^{6,1}$, A.C. Fernandes$^{6}$, J. Puibasset$^{3}$, and C. Oliveira$^{6}$}

\address{$^{1}$Centro de F\'isica Nuclear, Universidade de Lisboa, 1649--003 Lisbon, Portugal \\
$^{2}$Department of Physics, University of Chicago, Chicago IL, 60637 USA \\
$^{3}$Groupe de Physique des Solides (UMR CNRS 75--88), Universit\'e Paris 7 \& 6, 75251 Paris, France \\
$^{4}$Laboratoire Souterrain \`a bas Bruit, 84400 Rustrel--Pays d'Apt, France \\
$^{5}$Pacific Northwest National Laboratory, Richland, WA 99352   USA \\
$^{6}$Instituto Tecnol\'ogico e Nuclear, Estrada Nacional 10, 2686--953 Sacav\'em, Portugal \\
$^{7}$Department of Electronics, Instituto Superior T\'ecnico, Av. Rovisco Pais 1, 1049--001 Lisbon Portugal}

\maketitle

\abstracts{SIMPLE is an experimental search for evidence of spin-dependent dark matter, based on superheated droplet detectors using C$_{2}$ClF$_{5}$. We report preliminary results of a 0.6 kgdy exposure of five one liter devices, each containing $\sim$10 g active mass, in the 1500 mwe LSBB (Rustrel, France). In combination with improvements in detector sensitivity, the results exclude a WIMP--proton interaction above 5 pb at M$_{\chi}$ = 50 GeV/c$^{2}$.}

\section{Introduction}

SIMPLE\cite{jprl,jnjp} is is one of two experiments\cite{Hamel} to search for evidence of spin-dependent dark matter using fluorine--loaded superheated droplet detectors (SDDs). The detector is based on the nucleation of the gas phase by energy deposition in the superheated liquid. The conditions for bubble nucleation imply energy depositions of order $\sim$200 keV/$\mu$m, rendering SIMPLE SDDs effectively insensitive to most of the traditional backgrounds which plague the majority of the conventional dark matter search detectors.

In 2000, we reported\cite{jprl} the first exclusion limits from a prototype SDD with 9.2 g active mass operated for 16 days. Refrigerant-free 'dummy' modules however yielded signals identical to bubble nucleations arising from pressure microleaks in the SDD caps at a rate of 1 event per day\cite{jprl}, which comprised the mahority of the prototype signal.

We here report the preliminary results of a measurement using five detectors of similar construction, with design modifications in the capping of the detectors which result in dummy device rates a factor 10 less than previous, and an overall factor 2 reduction in our previous exclusion limits. Although still limited by statistics, the results clearly demonstrate the experiment competivity with the higher exposure search experiments.

\section{Experimental}

The detectors, ranging in active freon mass from 9.2--16.6 g with a total of $\sim$60 g, were installed in the GESA area of the LSBB. The SIMPLE SDDs, based on C$_{2}$ClF$_{5}$ (R-115), are fabricated inhouse with a 1--3\% loading, according to previously--described procedures\cite{jnjp}.

The set was placed inside a thermally--regulated 700 liter water bath, surrounded by three layers of sound and thermal insulation, resting on a dual vibration absorber. A hydrophone is placed within the detector water bath, and a second acoustic monitor positioned outside the shielding.

A bubble nucleation is accompanied by an acoustic shock wave, which is detected by a piezoelectric microphone embedded in a plastic finger (transducer) immersed in a glycerin layer at the top of the detector. The transducer signal was amplified a factor 10$^{5}$; in the case of an event in any of the detectors, the temperature, pressure, and threshold voltage for each device, plus its waveform trace and fast Fourier transform, were recorded in a LabView platform.

The operating pressure (2 atm) and temperature (9 $^{o}$C) were chosen to reduce the detector sensitivity to background, while preserving WIMP sensitivity. In these conditions, predominant background sources are either neutron or $\alpha$--particles, or electronics.

Due to their low stopping power, $\gamma$--rays below 6 MeV can produce background nucleations only through $\gamma$--induced high stopping power electrons (Auger electron cascades following interactions of environmental $\gamma$--rays with Cl atoms in the refrigerant\cite{jnjp}, visible above 15 $^{o}$C) or recoil nuclei from $\gamma$ scattering (kinematically below threshold for $T < 12 ^{o}$C).

The detector response to $\alpha$'s was described elsewhere\cite{jnjp}. The presence of a small ($\sim 10^{-4}$ pCi/g) $^{228}$Th contamination was measured via low--level $\alpha$ spectroscopy, yielding an overall background level of $<$ 0.5 events/kg freon/dy.

The response of smaller SDDs to various neutron fields has been studied extensively and found to match theoretical expectations\cite{harper,Lo}. The SIMPLE detector response to neutrons has been investigated using $^{252}$Cf and monochromatic low energy neutron beams generated by filtering the thermal column of the Portuguese Research Reactor\cite{jnjp,nima}. In both cases, the expected nucleation rate was calculated as a function of temperature via MCNP4 simulations of the response, following Refs. \refcite{Lo,jprd}. The $^{252}$Cf measurements yielded a detection efficiency of 34\% for 2 atm operation, consistent with the filter irradiations\cite{nima}, and in good agreement with the thermodynamic calculations.

The metastability of a superheated liquid is described by the homogeneous nucleation theory\cite{Seitz}, which predicts a spontaneous nucleation rate exponentially decreasing with decreasing temperature. This process should not be significant for rates down to the level of current measurement.

\section{Data analysis}

As shown in Table \ref{table1}, the detectors were operated at 9$^{o}$C  for 10 days, 3$^{o}$C  for 15 days. Additional experiments were performed at 14$^{o}$C in order to assess detector performance. The data record was filtered according to the criteria that (i) one and only one detector had a signal, and (ii) no monitoring detector had a signal.

\begin{table}[ph]
\tbl{Raw data results, without acoustic detection efficiency or background correction.}
{\footnotesize
\begin{tabular}{@{}cccccc@{}}
\hline
{} &{} &{} &{} &{}\\[-1.5ex]
run & exposure & T & P & anticoincidence & 5.5--6.5 kHz \\[1ex]
{} & (dy) & ($^{o}$C) & (bar) & (ev/kgdy) & (ev/kgdy)\\[1ex]
\hline
{} &{} &{} &{} &{}\\[-1.5ex]
2704 &10 &8.9 &2.0 & $197 \pm 18$ & $143 \pm 16$ \\[1ex]
2805 &14 &3.2 &1.9 & $64 \pm 9$ & $48 \pm 8$ \\[1ex]
\hline
\end{tabular}\label{table1} }
\vspace*{-13pt}
\end{table}

The fast Fourier transform of the transducer signal comprises a well--defined frequency response, with a primary harmonic at $\sim$6 kHz and a time span of a few milliseconds: only filtered events with a primary harmonic between 5.5--6.5 kHz were accepted.

The 3.2 $^{o}$C results, not being WIMP--relevant data, were used to estimate a lower limit on the overall background rate. Following the purification studies of Ref. \refcite{jnjp}, the 9 $^{o}$C--to--3 $^{o}$C rate ratio is $\sim$2.5, yielding a difference of 23$\pm$26 events/kgdy. At this level of exposure, the signal rate is almost entirely due to contributions from known background origins.

\begin{figure}[h]
\centerline{\epsfxsize=7cm\epsfbox{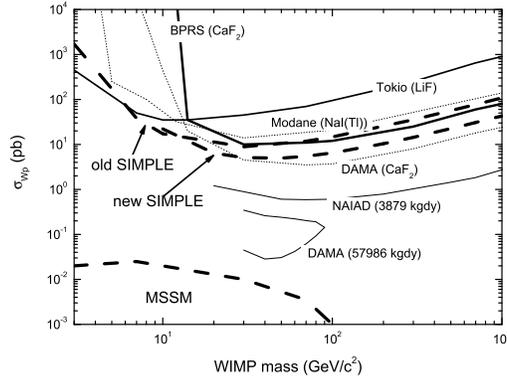}}   
\caption{comparison of preliminary present SIMPLE limits with those of Ref. 1; the DAMA/NaI and NAIAD have been updated. The WIMP characteristic velocity is 230 km/s, the earth velocity in May is 257 km/s, and $\rho$ = 0.3 GeVc$^{-2}$cm$^{-3}$. \label{sigps}}
\end{figure}

The cosmological parameters and method described in Ref. \refcite{lwsm} are used in the calculation of the WIMP elastic scattering rates. A comparison of present with previous results is shown in Fig. \ref{sigps}, indicating the factor 2 improvement over the prototype results. Note that simply assuming {\it no} WIMPs were detected would yield an order of magnitude better limit on the expected WIMP rate.

\begin{figure}[h]
\centerline{\epsfxsize=8cm\epsfbox{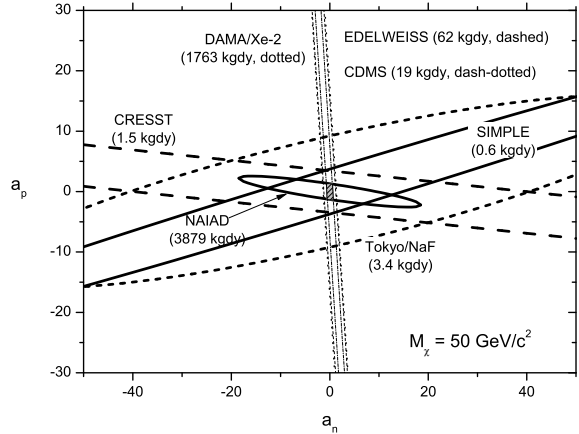}}   
\caption{a$_{p}$--a$_{n}$ plots for SIMPLE (0.6 kgdy), NAIAD (3879 kgdy), CRESST-I (1.51 kgdy), Tokyo/NaF (3.38 kgdy) and DAMA/Xe-2 (1763 kgdy) for WIMP mass of 50 GeV/c$^{2}$, as indicated. Also shown are the single nuclei spin-INdependent EDELWEISS (62 kgdy) and CDMS (19 kgdy). The unexcluded region of each experiment lies inside its respective contour.
\label{status}}
\end{figure}

Fig. \ref{sigps} is obtained within a model--dependent formulation. The data were also analyzed using a model--independent formalism\cite{fragiul}, in which the odd-group approximation for the WIMP-nucleus cross section $\sigma_{A}$ is abandoned and the full spin-dependent interaction with all isoptope nucleons is taken into account: $\sigma_{A} \propto (a_{p}<S_{p}>+a_{n}<S_{n}>)^{2}$, where $a_{p,n}$ ($S_{p,n}$) are the nucleon coupling strengths (spins) and $<S_{p}>$, $<S_{n}> \neq 0$, respectively. Since the phase space is now 3-dimensional, the results can be displayed by projection onto the $a_{p}$--$a_{n}$ plane for each given WIMP mass M$_{\chi}$, as shown in Fig. \ref{status} at 90\% C.L. for M$_{\chi}$=50 GeV/c$^{2}$ (which is in the DAMA/NaI--preferred range\cite{damanai}), together with the NAIAD\cite{naiad}, Tokyo/NaF\cite{naf}, CRESST-I\cite{cresst} and DAMA/Xe-2\cite{damaxe} experiments. Within this formulation, the SIMPLE results are already seen to eliminate a large part of the parameter space allowed by NAIAD at this mass cut. The CRESST result is assumed to be a result of Al only, since $^{16}$O is an even-even, spinless and doubly magic nucleus with no magnetic moment.

We also show the impact of the EDELWEISS\cite{EDELWEISS} and CDMS\cite{CDMS} experiments, customarily considered as spin-INdependent searches, which by themselves are surprisingly even more efficient in reducing the allowed parameter space of the NAIAD--DAMA/Xe-2 intersection.

The reason for the large impact of the fluorine--based detectors is (i) the relative sign of $<S_{n}>/<S_{p}>$ opposite to I, and (ii) both $<S_{n}>$ and $<S_{p}>$ are non--negligible. The near--orthogonality of the fluorine ellipses results from (i); chlorine and sodium, the other nuclei present in the detectors, do not fulfill condition (ii), and are essentially spectators for WIMP--neutron interaction detection. Nevertheless, they make the unexcluded regions of the experiments closed ellipses instead of open conics\cite{fragiul}, with the chlorine being the weaker of the two constraint--wise. This weakness is responsible for the high eccentricity of the SIMPLE ellipse, owing to the low spin values of $^{35}$Cl and low concentration of $^{37}$Cl.  

Despite the small active detector mass, the limits reflect the favourable $^{19}$F nuclear spins, and the reduced background inherent to a detection method essentially blind to the traditional backgrounds. Furthermore, the temperature--dependent threshold of the detector allows a background estimate at a temperature at which the detector is no longer sensitive to neutralino--induced events. The present SIMPLE limits remain constrained by the large statistical uncertainty associated with the short exposure accumulated so far, as well as continuing "anomalies" in the electronic response, both of which are currently being addressed.

\section*{Acknowledgements}
This work was supported by grant POCTI/FNU/43683/2002 of the Portuguese Foundation for Science and Technology (FCT), co--financed by FEDER.

\end{document}